# A proposed increase in retinal field-of-view may lead to spatial shifts in images

*Rupak Doshi[1,2*] and Philip J. R. Day[3]*

[1]Dept. of Pharmacology, University of Cambridge, Tennis Court Road, Cambridge CB2 1PD.

[2]St. Edmund's College, University of Cambridge, Mt. Peasant, Cambridge CB3 0BN.

[3]School of Translational Medicine, Manchester Interdisciplinary Biocentre, University of Manchester. 131, Princess Street Manchester M1 7DN. UK

[*]Corresponding Author: rd368@cam.ac.uk, Ph: +44 (01223) 334073, Fax: +44 (01223) 334100

**Abstract**

Visual information determines majority of our spatial behavior. The eye projects a 2-D image of the world on the retina. We demonstrate that when a monocular-like imaging system operates entirely with optically dense fluids, an increase in field-of-view (FOV) is observed compared to an experimental condition, where the ocular medium is optically neutral. Resulting spatial shifts in the retinal image towards the fovea complement the photoreceptor distribution pattern, incidentally revealing a new role for ocular fluids in the image space. Possible effects on the perceived egocentric object location are discussed.

**Introduction**

*"Optimal visual performance depends on the maintenance of a high-quality optical image on the retina"* – Neil Charman [1]

*"...we need to ask at the outset how good a replica of the outside world the retinal image really is."* - Gerald Westheimer [2]

The above-mentioned quotes summarize our present knowledge, according to which, pictorial representations of the world which are permitted by refraction of light at the air-eye boundary, are sole and crucial substrates for vision perception. The initiation of the sensory-response process, i.e. the optics of image formation and retinal image space, has been well-studied [2-4]. Various aberrations pertaining to the optical processes in the human eye have been described in considerable detail [2, 3, 5, 6]. While the lens and cornea have attracted much attention in this regard [7], the major known optical effect associated with the ocular fluids is with respect to the absorption of incoming light [8, 9]. In this brief letter, using a simple bench-top model eye, we report an observation which, we propose, is a previously undocumented role of ocular fluids pertaining to the retinal image space.

It is well-known that human eyes have evolved in way where foveal vision's accuracy surpasses that in the periphery [10]. This is explained by the dominance of cones in the fovea, and by the ocular imaging process where the maximum amount light hits the optical centre [2, 11]. Recently, it has also been suggested that architectural aspects of retinal physiology may be evolutionary tricks to conserve space and energy [12]. Our proposition supports this idea and indicates another advantageous reason for the development of foveal preference.

**Hypothesis**

The major principles of geometric optics guiding the eye and a camera are similar [6]. However two major differences are: 1. a roughly convex shaped cornea is the first and primary refracting element in the eye, whereas it is a biconvex lens in the camera (Figure 1a, c), and 2. the retinal image formation process is entirely maintained at a reduced speed of light due to the continuous layers of optically dense ocular elements that populate the space between the first refracting boundary (cornea) and the photographic film (retina), which in the case of cameras, is occupied by air [13, 14] (Figure 1c).

The presence of ocular fluids in the image space of the eye, and water in the object space for a camera, should be expected to have opposing effects. This is because of the difference in shapes of the cornea and lens; while both have their convex surfaces facing the object, a biconvex lens has the additional convex surface facing away from the former (Figure 1a, c). Known is that cameras immersed in water, suffer a decrease in field-of-view (FOV) due to the presence of a medium exceeding the optical density of air, in the object's surrounding space (R.I.'s of air and water are ~ 1.00 and 1.33 respectively) [15, 16] (Figure 1a, b). Conversely, when an optically dense medium dominates the image space in the eye, an opposing effect should be expected, i.e. an increase in FOV (Figure 1c, d).

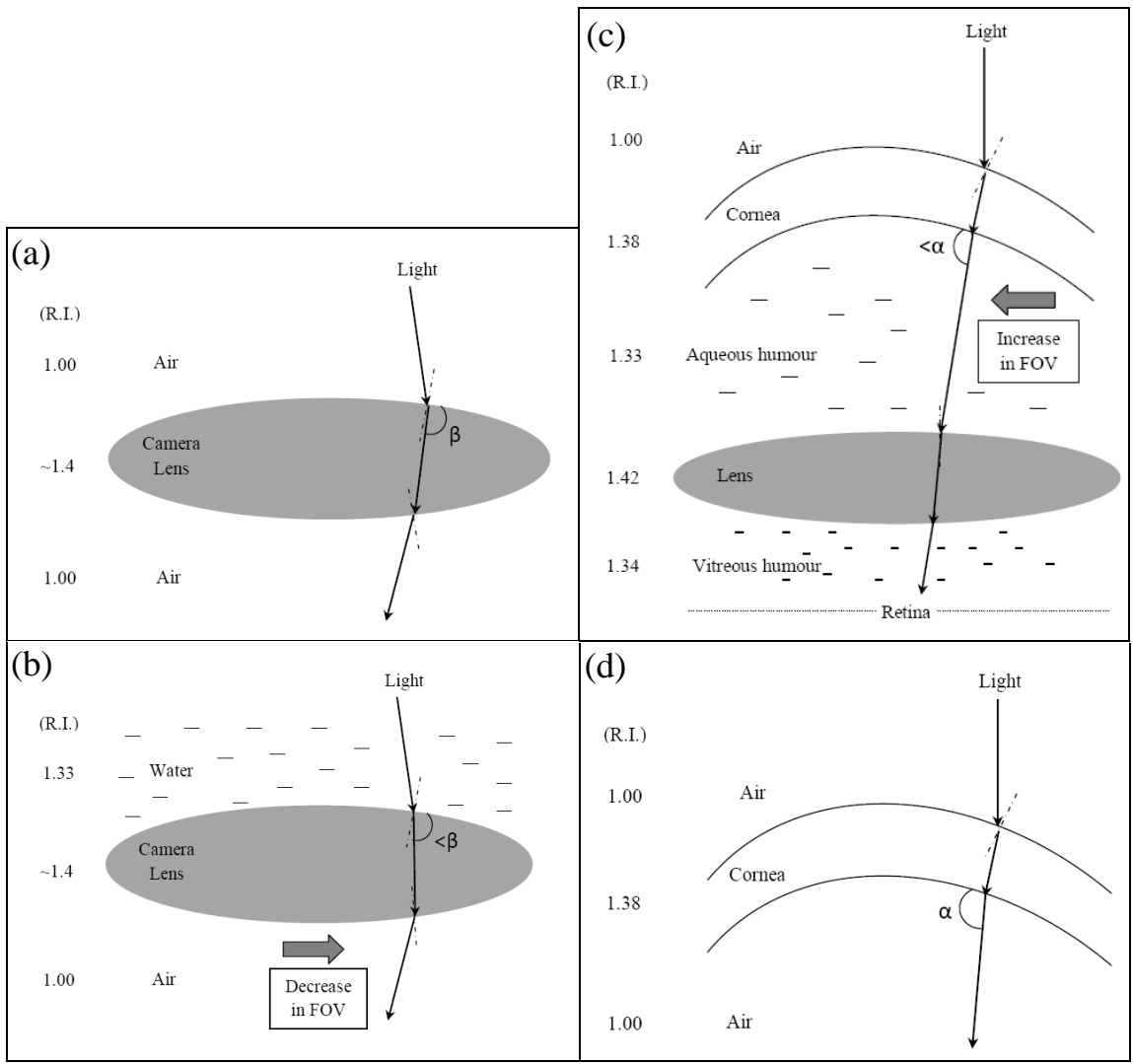

**Figure 1**: Working hypothesis. (a), (b) A camera (assumed R.I. of the biconvex lens ~1.4) underwater experiences a decrease in its field-of-view (FOV) due to water in the object space, as shown by the direction of arrow and change in angle β. (c), (d) In contrast, an eye's FOV increases due to the presence of ocular fluids in the image space, as shown by the arrow direction and angle α. (1c adapted from [13]). Figures not drawn to scale, approximate refractive indices (R.I.) are shown alongside.

**An increase in retinal FOV with concomitant alterations in spatial coordinates of the image**

We have demonstrated the increase in FOV using a bench-top human eye model (15 cm x 17 cm x 10 cm high) in which, water is modeled for the aqueous and vitreous fluids and thus enables a direct comparison between images obtained on the retina with or without water in the image space [17] (Figure. 2a). Similar, water-filled model eyes have been used for applications such as characterizing intraocular lens implants [18].

As is evident in Figure 2 b-d, the under-water focused image of the LED light source was "translocated" in a direction towards the optical centre (see arrow in Figure 2c), and an increase in FOV was noted as the retinal image had moved from partially off-scale to fully on-scale (see also Table 1a). This shift was confirmed as not being due to the increasing power of the lens (data not shown), but due to the presence of water. We further characterized the magnitude of spatial shift's dependence on the horizontal retinal image location (i.e. eccentricity, while maintaining the same vertical object distance from the cornea, i.e. ~50 cm). Our observations revealed that, 1. the spatial differences in the image with or without water, were more pronounced at the periphery of the retina, decreasing progressively towards the optical centre (Table 1 a-d) and, 2. the images converged towards the centre from both left and right (relative to the reader looking at a photograph of the retinal image) halves of the retina (Table 1 a-d and e-h).

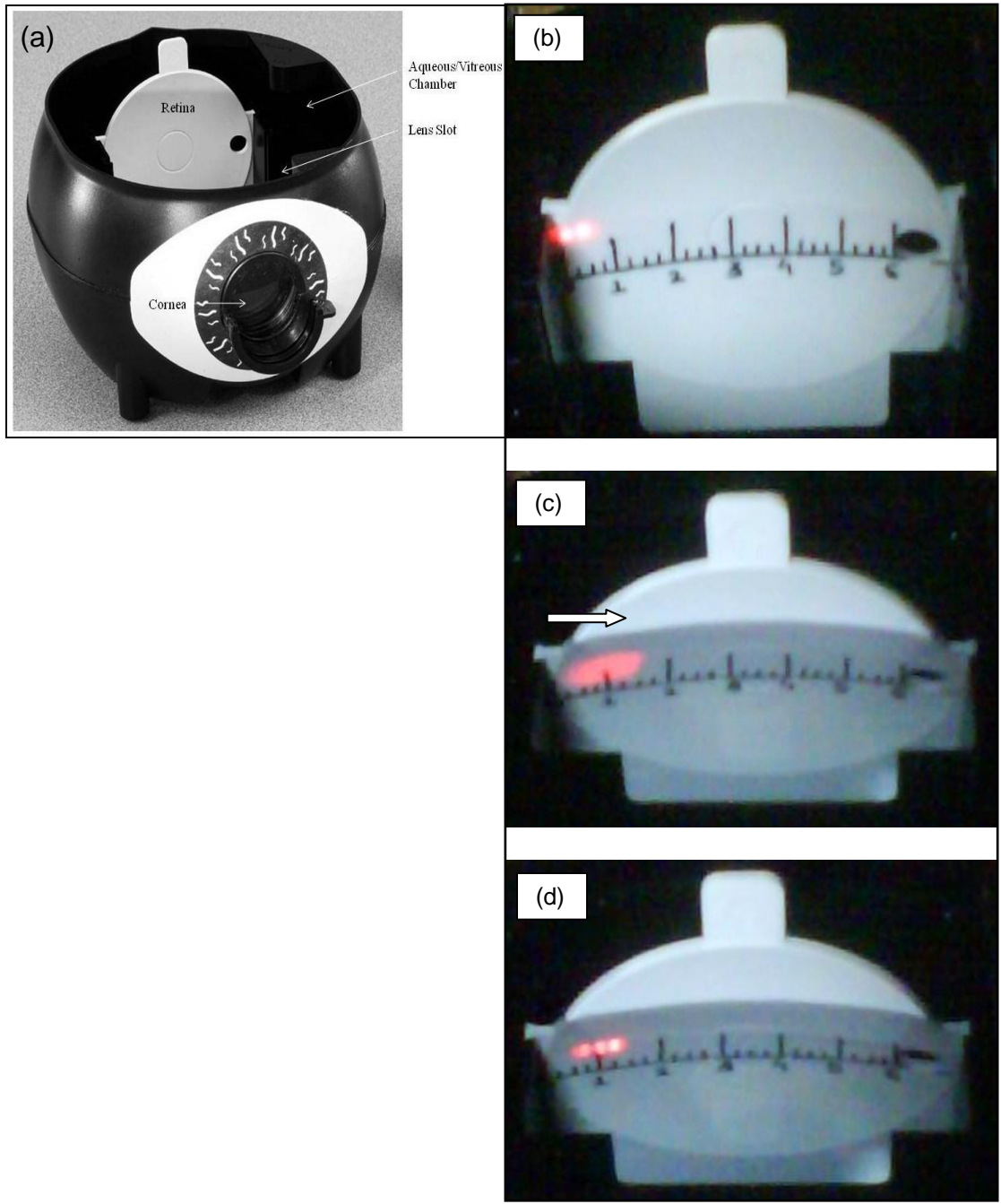

**Figure 2:** Alteration in the 2-D spatial coordinates of images on the retina is due to an increased FOV in the presence of an optically dense medium. (a) The human eye model used in this study employs water (R.I. ~1.33) to model for the aqueous and vitreous fluids (Figure adapted from [17]). Image of an LED light source placed at ~50 cm from the cornea of the eye model, obtained on the retina using (b) a +2.5d power lens (+ 400 mm focal length in air) in air, (c) after the addition of water to the aqueous/vitreous chamber till the cornea and lens were completely submerged, and (d) with a +8.3d power (+120 mm focal length in air) lens added to increase focus in the presence of water. The retinal scale has arbitrary units and has been drawn as a visual guide.

| No. | Retinal Location-Air (arbitrary units) | Retinal Location-Water (arbitrary units) |
| --- | --- | --- |
| a. | 0.5 (off-scale) (Fig. 2b) | 1.5 (on-scale) (Fig. 2d) |
| b. | 1.5 | 2.0 |
| c. | 2.5 | 2.75 |
| d. | 3.5 | 3.5 |
| e. | 4.5 | 4.25 |
| f. | 5.5 | 5.0 |
| g. | 6.5 | 6.0 |
| h. | ~7.5 (off-scale) | 5.75 (on-scale) |

**Table 1**: Locations of the right-hand (of the reader) edge of the image of the LED light source on the retinal scale, with conditions as per Figure 2b (retinal location-air) and Figure 2d (retinal location-water). The light source was displaced along a horizontal arc relative to the vertical axis of the eye model in order to get the desired "air" retinal location (i.e. object distance was maintained at ~50 cm in all cases, changing only the eccentricity of the image).

These data were strong indications that spatial shifts in the retinal image were due to refraction pattern alterations in the presence of water (see Figure 1c, d). Interestingly, the earliest technique to simulate a wide-angle view was documented in 1905, where the body of a pinhole camera was filled with water [19]. Our observations suggest that the human eye behaves like a water-filled *camera obscura* in this regard, having its FOV substantially increased, strengthening the historical comparisons between the two [6]. Notably, we could also obtain similar results using exclusive corneal focusing conditions, regardless of the use of a modeled-pupil (the only difference observed with the use of a pupil was the decrease in intensity of the image), indicating that the alterations were independent of additional lenses or geometric area of the lens used to focus.

We acknowledge that this model eye might not be as accurate to provide absolute quantitative information. However it suffices the requirements of this experiment, and the relative patterns observed here could be expected to be reproduced in the physiological monocular eye. Interestingly, the eye uses small changes in refractive indices to bring displaced images into focus continuously through accommodation [4], and many optical aberration corrections also employ moving images on the retina through refractive changes. Thus, it would not be surprising that in its pristine state the eye would already use the presence of optical fluids to move retinal images towards the fovea.

**Proposal**

We have shown that for the same object, an image formed by the eye with its ocular fluids, is spatially different from that compared to one without. The only retinal location which would not experience this difference is the optical centre of the imaging system. The human eye is not a centered system, with the foveal line of sight being tilted from the horizontal optical axis [2]. Consequently, all focused images are expected to be subjected to a certain degree of the spatial shift discussed here, with more drastic effects at the periphery.

It was recently suggested that retinotopic input has profound influences during development of the visual system [20]. Thus, our observations may reveal another reason why human eyes might have evolved to employ foveal vision to a much larger extent. Likely, is that the increased FOV is a space-saving strategy supporting the recent suggestion [12]. Studies on virtual environments have suggested that increasing FOV leads to an elevation in the sense of presence ("sensation of reality") [21, 22] and better performance in hand-eye coordination tasks [23, 24]. A large FOV is likely to offer an evolutionary advantage; such a mechanism is maintained from fishes to all higher mammals. Ocular fluids also serve vital functions of circulating metabolic needs and maintaining intro-ocular pressure [25]. Thus being highly advantageous, it makes for good logic that the eyes have evolved to utilize the increased FOV in an effective manner.

**Possible effects on vision perception**

*"…if one scrambles (or alters) the spatial arrangement of the (visual) image at a fine scale any realistic chance of reconstructing the original is lost." [26]*

Although almost all known optical phenomena of the eye can be explained by geometric, first order Gaussian, and wave optics, a detailed reproduction of what the eye actually sees has been hindered, due to unavailable technology that can capture a high-resolution image of the functionally live retina [2, 5, 27]. So the question, whether normal retinal images are faithful snapshots of the real world in 2-D, remains to be incontrovertibly answered. In this regard, our observations may raise questions regarding the veridicality of egocentric object locations.

The physical space-time dimensions are well-known to be dependent on the speed of light [28]. In the eye, both the phase and group velocities of light are hindered [29]. Based on what is known about relativistic space distortions, the spatial shifts in this letter do not fall in the same category. Nevertheless, they do result due to the presence of optically dense media in the image space, which in turn is responsible for maintaining a reduced light speed throughout the imaging process [14, 29].

In recent years, accurate maps of the visual field have been identified in cortical regions, which have been used to explain the high-resolution spatial property of vision [26]. These represented space maps in 12 visual cortical areas were shown to be retinotopic not spatiotopic [30]. Although internal optic [7] and neural mechanisms [31] exist to account for many optical errors of vision, at present it is unclear if such an ocular or perceptual process exists, that compensates for the retinal image shifts discussed here. Additionally, the neural processes would not have a calibrative reference-frame as vision itself is known to be the best resolved spatial sense [32, 33]. However prior to the above investigation, whether the said spatial shifts remain after binocular integration, needs to be studied.

If the spatial alterations in retinal images described in this letter, would bear impressions in the related perceived visual space, the consequences could be far-reaching. Retinal and extra-retinal factors affecting ego- and exo-centric object locations have been thoroughly studied [34-36], and this would provide an added retinal factor.

A recent study concluded that errors in visually directed rapid pointing tasks increase, on increasing the distance of images from the foveal centre [37]. We have shown that the non-veridicality of retinal images might increase towards the retinal periphery (Table 1). Whether these pointing errors are due to retinal displacement, presents an interesting question.

**Conclusion**

A detailed pixel-wise retinal image still eludes vision science. Our demonstration and resulting proposal, warrants detailed analysis of the retinal image and investigation of whether retinal space, and (or) perceived visual space, is an altered version of veridical space. In addition to having important implications for visual psychology and philosophy, our proposal may also assist in developing a novel demarcation for visual proximal and distal stimuli (by Gestalt terminology) [38], i.e. objects and their retinal representations are maintained at different speeds of light.

**Funding**

The human eye model was purchased with the help of a generous award from St. Edmund's College, University of Cambridge, UK.

**Competing Interests Statement**

The authors declare no competing interests.

**Acknowledgements**

We would like to thank Prof. Paul V. McGraw, School of Psychology, University of Nottingham for critical reading and comments on the manuscript, and Prof. Nicholas Wade, Dept. of Psychology, University of Dundee, and Prof. David Whitaker, School of Life Sciences/Vision Sciences, University of Bradford for their valuable advice.

**References**

1.  N. Charman, "http://www.ls.manchester.ac.uk/research/researchgroups/eyeandvisionsciences/people/index.aspx?PersonID=384&view=research."
2.  G. Westheimer, "Specifying and controlling the optical image on the human retina," Progress in Retinal and Eye Research **25**, 19-42 (2006).

3.	J. A. M. Jennings, and W. N. Charman, "Off-axis image quality in the human eye," Vision Research **21**, 445-455 (1981).
4.	A. Gullstrand, "How I found the mechanism of intracapsular accommodation," Nobel Lecture, http://nobelprize.org/nobel_prizes/medicine/laureates/1911/gullstrand-lecture.pdf (1911).
5.	N. J. Wade, "Image, eye, and retina (invited review)," J. Opt. Soc. Am. A **24**, 1229-1249 (2007).
6.	N. J. Wade, and S. Finger, "The eye as an optical instrument: from camera obscura to Helmholtz's perspective," Perception **30**, 1157-1177 (2001).
7.	P. Artal, A. Guirao, E. Berrio, and D. R. Williams, "Compensation of corneal aberrations by the internal optics in the human eye," Journal of Vision **1**, 1-8 (2001).
8.	T. J. T. P. van den Berg, and H. Spekreijse, "Near infrared light absorption in the human eye media," Vision Research **37**, 249-253 (1997).
9.	R. A. Weale, "Spectral Sensitivity Curves and the Absorption of Light by the Ocular Media," British Journal of Ophthalmology **37**, 148-156 (1953).
10.	S. J. Anderson, K. T. Mullen, and R. F. Hess, "Human peripheral spatial resolution for achromatic and chromatic stimuli: limits imposed by optical and retinal factors," The Journal of Physiology **442**, 47-64 (1991).
11.	A. C. Christine, R. S. Kenneth, E. K. Robert, and E. H. Anita, "Human photoreceptor topography," The Journal of Comparative Neurology **292**, 497-523 (1990).
12.	V. Balasubramanian, and P. Sterling, "Receptive fields and functional architecture in the retina," The Journal of Physiology **587**, 2753-2767 (2009).
13.	M. Hollins, *Medical Physics* (Neilson Thornes, Bath Advanced Science, Cheltenham, 2001).
14.	M. B. James, and D. J. Griffiths, "Why the speed of light is reduced in a transparent medium," American Journal of Physics **60**, 309-313 (1992).
15.	S. F. Ray, *Applied Photographic Optics: Lenses and Optical Systems for Photography, Film, Video, Electronic and Digital Imaging* (Focal Press, Oxford, 2002).
16.	N. Wu, *How to photograph under-water* (Stackpole Books, Mechanicsburg, 1994).
17.	U. S. A. PASCO Scientific, "http://store.pasco.com/pascostore/showdetl.cfm?&DID=9&Product_ID=56044&Detail=1."
18.	R. J. Landry, I. K. Ilev, T. J. Pfefer, M. Wolffe, and J. J. Alpar, "Characterizing reflections from intraocular lens implants," Eye **21**, 1083-1086 (2007).
19.	J. M. Franke, "Field-widened pinhole camera," Applied Optics **18**, 2913-2914 (1979).
20.	R. Rajimehr, and R. B. H. Tootell, "Does Retinotopy Influence Cortical Folding in Primate Visual Cortex?," Journal of Neuroscience **29**, 11149-11152 (2009).
21.	T. Hatada, H. Sakata, and H. Kusaka, "Psychophysical analysis of the "sensation of reality" induced by a visual wide-field display," SMPTE Journal **89**, 560-569 (1980).
22.	J. D. Prothero, and H. G. Hoffman, "Widening the Field-of-View Increases the Sense of Presence in Immersive Virtual Environments," HITLab Technical Report R-95-5, Available online: http://www.hitl.washington.edu/publications/r-95-95/ (1995).

23. K. Arthur, "Effects of field of view on task performance with head-mounted displays," in *Conference companion on Human factors in computing systems: common ground*(ACM, Vancouver, British Columbia, Canada, 1996).
24. K. W. Arthur, "Effects of Field of View on Performance with Head-Mounted Displays," in *PhD Thesis, Department of Computer Science*(University of North Carolina at Chapel Hill, Chapel Hill, 2000).
25. D. V. N. Reddy, and V. E. Kinsey, "Composition of the Vitreous Humor in Relation to That of Plasma and Aqueous Humors," Archives of Ophthalmology **63**, 715-720 (1960).
26. B. A. Wandell, S. O. Dumoulin, and A. A. Brewer, "Visual Field Maps in Human Cortex," Neuron **56**, 366-383 (2007).
27. L. N. Thibos, "Formation and sampling of the retinal image," in *Seeing*, K. K. D. Valois, ed. (London: Academic press, 2000), pp. 1-54.
28. E. F. Taylor, and J. A. Wheeler, *Spacetime Physics* (W. H. Freeman and Company, 1966).
29. W. Drexler, C. K. Hitzenberger, A. Baumgartner, O. Findl, H. Sattmann, and A. F. Fercher, "Investigation of Dispersion Effects in Ocular Media by Multiple Wavelength Partial Coherence Interferometry," Experimental Eye Research **66**, 25-33 (1998).
30. J. L. Gardner, E. P. Merriam, J. A. Movshon, and D. J. Heeger, "Maps of Visual Space in Human Occipital Cortex Are Retinotopic, Not Spatiotopic," Journal of Neuroscience **28**, 3988-3999 (2008).
31. P. Artal, L. Chen, E. J. Fernàndez, B. Singer, S. Manzanera, and D. R. Williams, "Neural compensation for the eye's optical aberrations," Journal of Vision **4**, 281-287 (2004).
32. S. Shimojo, and L. Shams, "Sensory modalities are not separate modalities: plasticity and interactions," Current Opinion in Neurobiology **11**, 505-509 (2001).
33. C. Spence, and J. Driver, *Crossmodal Space and Crossmodal Attention* (Oxford University Press, 2004).
34. R. F. Hess, and A. Hayes, "The coding of spatial position by the human visual system: Effects of spatial scale and retinal eccentricity," Vision Research **34**, 625-643 (1994).
35. P. Magne, and Y. Coello, "Retinal and extra-retinal contribution to position coding," Behavioural Brain Research **136**, 277-287 (2002).
36. J. M. Loomis, J. A. D. Silva, J. W. Philbeck, and S. S. Fukusima, "Visual Perception of Location and Distance," Current Directions in Psychological Science **5**, 72-77 (1996).
37. A. Ma-Wyatt, and S. P. McKee, "Initial visual information determines endpoint precision for rapid pointing," Vision Research **46**, 4675-4683 (2006).
38. K. Koffka, *Principles of Gestalt Psychology* (Harcourt, Brace, New York, 1935).